\PassOptionsToPackage{table,xcdraw}{xcolor}
\documentclass[9pt,conference]{IEEEtran}
\IEEEoverridecommandlockouts
\usepackage{cite}
\usepackage{amsmath,amssymb,amsfonts}
\usepackage{algorithmic}
\usepackage{multirow}
\usepackage{graphicx}
\usepackage{textcomp}
\usepackage{tikz}
\usepackage{color}
\usepackage{listings}
\usepackage{makecell}
\usepackage{float}
\usepackage{graphicx}
\newcommand*\emptycirc[1][1ex]{\tikz\draw (0,0) circle (#1);} 

\newcommand*\fullcirc[1][1ex]{\tikz\fill (0,0) circle (#1);}
\usepackage[ruled,vlined]{algorithm2e}
\usepackage{amsmath} 
\usepackage{bm} 
\usepackage{subcaption}
\definecolor{GrayCodeBlock}{RGB}{241,241,241}
\definecolor{BlackText}{RGB}{0,0,0}
\definecolor{RedText}{RGB}{0,158,96}
\definecolor{BlueText}{RGB}{0,0,255}
\definecolor{DahongText}{RGB}{255,0,0}
\definecolor{RedTypename}{RGB}{182,86,17}
\definecolor{GreenString}{RGB}{96,172,57}
\definecolor{PurpleKeyword}{RGB}{184,84,212}
\definecolor{GrayComment}{RGB}{170,170,170}
\definecolor{GoldDocumentation}{RGB}{180,165,45}

\lstdefinelanguage{rust}
{
    columns=fullflexible,
    keepspaces=true,
    frame=single,
    framesep=0pt,
    framerule=0pt,
    framexleftmargin=4pt,
    framexrightmargin=4pt,
    framextopmargin=5pt,
    framexbottommargin=3pt,
    xleftmargin=4pt,
    xrightmargin=4pt,
    backgroundcolor=\color{GrayCodeBlock},
    basicstyle=\footnotesize\color{BlackText},
    numbers=left,
    numberstyle=\tiny, 
    keywords={
        true,false,
        unsafe,async,await,move,
        use,pub,crate,super,self,mod,
        struct,enum,fn,const,static,let,mut,ref,type,impl,dyn,trait,where,as,
        break,continue,if,else,while,for,loop,match,return,yield
    },
    keywordstyle=\color{RedTypename},
    ndkeywords={
        bool,u8,u16,u32,u64,u128,i8,i16,i32,i64,i128,char,str,
        Self,Option,Some,None,Result,Ok,Err,String,Box,Vec,Rc,Arc,Cell,RefCell,HashMap,BTreeMap,
        macro_rules
    },
    ndkeywordstyle=\color{RedTypename},
    comment=[l][\color{GrayComment}\slshape]{//},
    morecomment=[s][\color{GrayComment}\slshape]{/*}{*/},
    morecomment=[l][\color{GoldDocumentation}\slshape]{///},
    morecomment=[s][\color{GoldDocumentation}\slshape]{/*!}{*/},
    morecomment=[l][\color{GoldDocumentation}\slshape]{//!},
    morecomment=[l][\color{RedText}\slshape]{\#},
    morecomment=[l][\color{BlueText}\slshape]{\...},
    morecomment=[l][\color{DahongText}\slshape]{\*},
    morecomment=[s][\color{RedTypename}]{\#![}{]},
    morecomment=[s][\color{RedTypename}]{\#[}{]},
    stringstyle=\color{GreenString},
    string=[b]"
}

\def\BibTeX{{\rm B\kern-.05em{\sc i\kern-.025em b}\kern-.08em
    T\kern-.1667em\lower.7ex\hbox{E}\kern-.125emX}}

\title{Unlocking a New Rust Programming Experience: Fast and Slow Thinking with LLMs to Conquer Undefined Behaviors\vspace{-10pt}
}

\author{
\IEEEauthorblockN{Renshuang Jiang}
\IEEEauthorblockA{\textit{National University of Defense Technology}\\
Changsha, China \\
rshuang@nudt.edu.cn}
\and
\IEEEauthorblockN{Pan Dong}
\IEEEauthorblockA{\textit{National University of Defense Technology}\\
Changsha, China \\
pandong@nudt.edu.cn}
\and
\IEEEauthorblockN{ Zhenlin Duan}
\IEEEauthorblockA{\textit{National University of Defense Technology}\\
Changsha, China \\
15779658249@163.com}
\and
\IEEEauthorblockN{ Yu Shi}
\IEEEauthorblockA{\textit{National University of Defense Technology}\\
Changsha, China \\
shiyu19@nudt.edu.cn}
\and
\IEEEauthorblockN{Xiaoxiang Fang}
\IEEEauthorblockA{\textit{National University of Defense Technology}\\
Changsha, China \\
xiaoxiang200@163.com}
\and
\IEEEauthorblockN{Yan Ding}
\IEEEauthorblockA{\textit{National University of Defense Technology}\\
Changsha, China \\
yanding@nudt.edu.cn}
\and
\IEEEauthorblockN{Jun Ma}
\IEEEauthorblockA{\textit{National University of Defense Technology}\\
Changsha, China \\
majun@nudt.edu.cn}
\and
\IEEEauthorblockN{Shuai Zhao}
\IEEEauthorblockA{\textit{Sun Yat-sen University}\\
Guangzhou, China \\
zhaosh56@mail.sysu.edu.cn}
\and
\IEEEauthorblockN{Zhe Jiang\textsuperscript{*}}
\IEEEauthorblockA{\textit{Southeast University}\\
Nanjing, China \\
zhejiang.uk@gmail.com}
}

\begin{document}

\maketitle

\begin{abstract}

To provide flexibility and low-level interaction capabilities, the ``unsafe'' tag in Rust is essential in many projects, but undermines memory safety and introduces Undefined Behaviors (UBs) that reduce safety.
Eliminating these UBs requires a deep understanding of Rust's safety rules and strong typing. 
Traditional methods require depth analysis of code, which is laborious and depends on knowledge design. 
The powerful semantic understanding capabilities of LLM offer new opportunities to solve this problem. 
Although existing large model debugging frameworks excel in semantic tasks, limited by fixed processes and lack adaptive and dynamic adjustment capabilities. 
Inspired by the dual process theory of decision-making (``Fast and Slow Thinking''), we present a LLM-based framework called RustBrain that automatically and flexibly minimizes UBs in Rust projects. 
Fast thinking extracts features to generate solutions, while slow thinking decomposes, verifies, and generalizes them abstractly. 
To apply verification and generalization results to solution generation, enabling dynamic adjustments and precise outputs, RustBrain integrates two thinking through a feedback mechanism. 
Experimental results on Miri dataset show a 94.3\% \textit{pass} rate and 80.4\% \textit{execution} rate, improving flexibility and Rust projects safety. 

\end{abstract}


\section{\vspace{-2.5pt}\textbf{Introduction}\vspace{-2.5pt}}

Rust, a systems programming language, is popular for its guarantees of memory safety and concurrency~\cite{b1,b2}. 
For flexibility, Rust has a secondary language hidden inside it which does not enforce memory safety guarantees: it’s called Unsafe Rust and works just like regular Rust (it’s called Safe Rust). 
Safe Rust ensures safety through strict compiler checks. Unsafe Rust uses \textit{unsafe} keyword to grant direct control over memory and system resources, bypassing certain compiler checks, with safety guarantees that depend on the developer. 
Therefore, this capability introduces the potential for Undefined Behaviors (UBs), a class of run-time bugs that are difficult to detect and can lead to severe security vulnerabilities and instability~\cite{b3,b4}. 
Since Unsafe Rust cannot be avoided~\cite{b5}, especially in system software, how to perform fixes is a key in ensuring Rust's safety. 

The main methods for fixing UBs in Unsafe Rust include safe substitution, inserting assertions, and semantic modification~\cite{b6,b7,b8,b9}. 
Rust's strongly typed system, along with its strict ownership, lifetime, and borrowing rules, makes this process challenging, requiring an in-depth understanding of complex code semantics~\cite{b40}. 
Although the types of Unsafe Rust operations are finite (five types of unsafe operation~\cite{b1,b5}), minor semantic variation can significantly affect how code modification is approached~\cite{b3,b9}, leading to situations where similar UB may not yield comparable solutions. 
Therefore, understanding semantics and possessing specialized knowledge are essential to addressing Rust safety issues. 
Traditional methods~\cite{b33,b34,b35,b36,b37} require engineers to apply specialized knowledge along with strict logical reasoning, judgment, and verification processes, where the efficiency of resolution is closely tied to experience and expertise. 
The powerful self-reasoning and semantic understanding capabilities of large language models (LLMs) offer new opportunities to enhance the safety of Rust programs~\cite{b10,b11,b12,b13}. 
Due to high training costs and a lack of high-quality data in specific domains, conventional LLMs struggle with deep understanding and complex relational reasoning when handling specialized languages like Rust. 
In contrast, existing large model repair frameworks~\cite{b12,b13}, while providing assistance for such problems, typically follow pre-designed and fixed repair processes, lacking the professional capabilities to tailor solutions based on the specific characteristics of the code. 

Inspired by human cognitive psychology, \textit{``Thinking Fast and Slow"}~\cite{b14}, we observe that LLMs provide a broad, generalized understanding that resembles the ``fast thinking'' of humans.
This thinking is rapid and intuitive, but it is also error-prone reasoning. 
Yet, unlike human cognition, LLMs lack a ``slow thinking'' mechanism, employing specialized knowledge and logical reasoning to enhance accuracy, identify discrepancies, and mitigate errors. 
This is crucial when addressing complex problems that demand deliberate and systematic approaches to detect inaccuracies and ensure correctness.
These two complementary processes -- fast, intuitive reasoning and slow, analytical reasoning -- form a holistic framework for understanding and solving complex problems, e.g., solving UBs in Rust.



\begin{figure}[t]
\centerline{\includegraphics[width=0.5\textwidth, trim=0.5cm 20.9cm 17.7cm 1cm, clip]{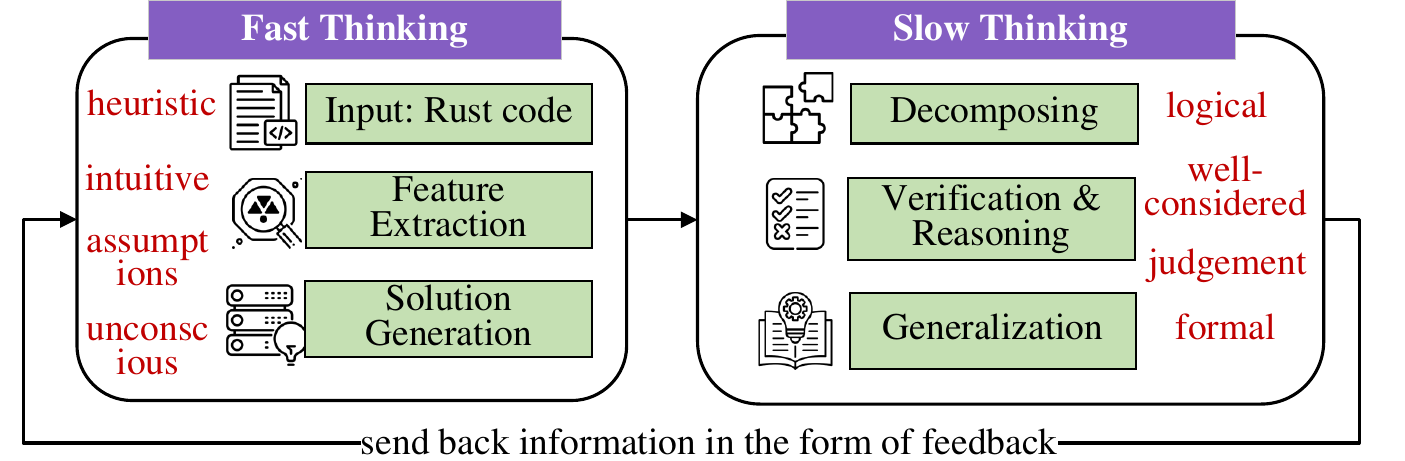}}
\vspace{-5pt}
\caption{RustBrain: Fast and Slow thinking LLM processes. }
\vspace{-17pt}
\label{overview}
\end{figure}


In light of this, we introduce RustBrain, a framework that integrates these cognitive processes, leveraging their respective strengths to autonomously and flexibly minimize UBs.
As depicted in Fig.~\ref{overview}, 
in the fast-thinking process, it extracts code information, identifies error types, and analyzes them using a broad knowledge base to generate multiple repair solutions. 
In the slow-thinking process, the framework in-depth analyzes and validates the repair solutions based on the four design principles: decomposition, verification, reasoning and generalization.
First, the solutions are decomposed and executed using multiple agents. 
In this stage, based on observations of Rust, we designed several agents,including three types of error-fixing agents, adaptive rollback agent, and abstract reasoning agent, to support better decomposition and verification. 
Then, the solution is fine-tuned based on verification results in slow-thinking, further optimizing the solution. 
Finally, to enhance the framework's generalization capability and better support precise solutions for similar errors, RustBrain establishes an evaluation and self-learning mechanism. 
This feeds the execution and evaluation results from the slow-thinking stage back into the fast-thinking, continuously optimizing solution generation and refining decision-making.

To evaluate RustBrain, we utilize a dataset containing UBs provided by Miri repository~\cite{b17}. 
The results demonstrate that it achieves 94\% \textit{pass} rate and 80\% execution semantic acceptability rate. 
Compared to the state-of-the-art Rust fix tool (RustAssistant~\cite{b13}), it \textit{pass} rate has increased by 30\% . 
Additionally, compared to engineer experts, the speed evaluation shows an improvement of up to 18x. 

\section{\textbf{RustBrain: An Overview}}
\subsection{Scope}
Our goal is to build a toolchain that flexibly and systematically evaluates the capabilities of RustBrain for eliminating UBs in Rust at compile time. 
Repaired code must pass the Miri compiler while preserving its expected semantics. 
Miri~\cite{b17} is a UB detection tool for Rust. 
Since many Rust code repairs are closely tied to context, the acceptability of semantics before and after the repair may vary significantly. 
Therefore, this paper validates semantics using test benchmarks composed of developer-repaired code.
Compared to direct testing through Miri, this semantic judgment approach may introduce a degree of subjectivity. This is due to factors such as the size of the benchmark (i.e., the probability of covering all possible semantics) and other variables.
Besides, we use Miri to detect Rust UBs and only considers Rust source files (i.e, files with .rs extension). 
Errors not detected by Miri are beyond the scope of this paper. 
In addition, RustBrain uses LLMs to guide the reduction of UBs, and we temporarily disregard Rust code that exceeds LLM tokens limits.

\subsection{RustBrain Overview}
\begin{figure}[t]
\centerline{\includegraphics[width=0.5\textwidth, trim=0.3cm 11cm 18cm 1cm, clip]{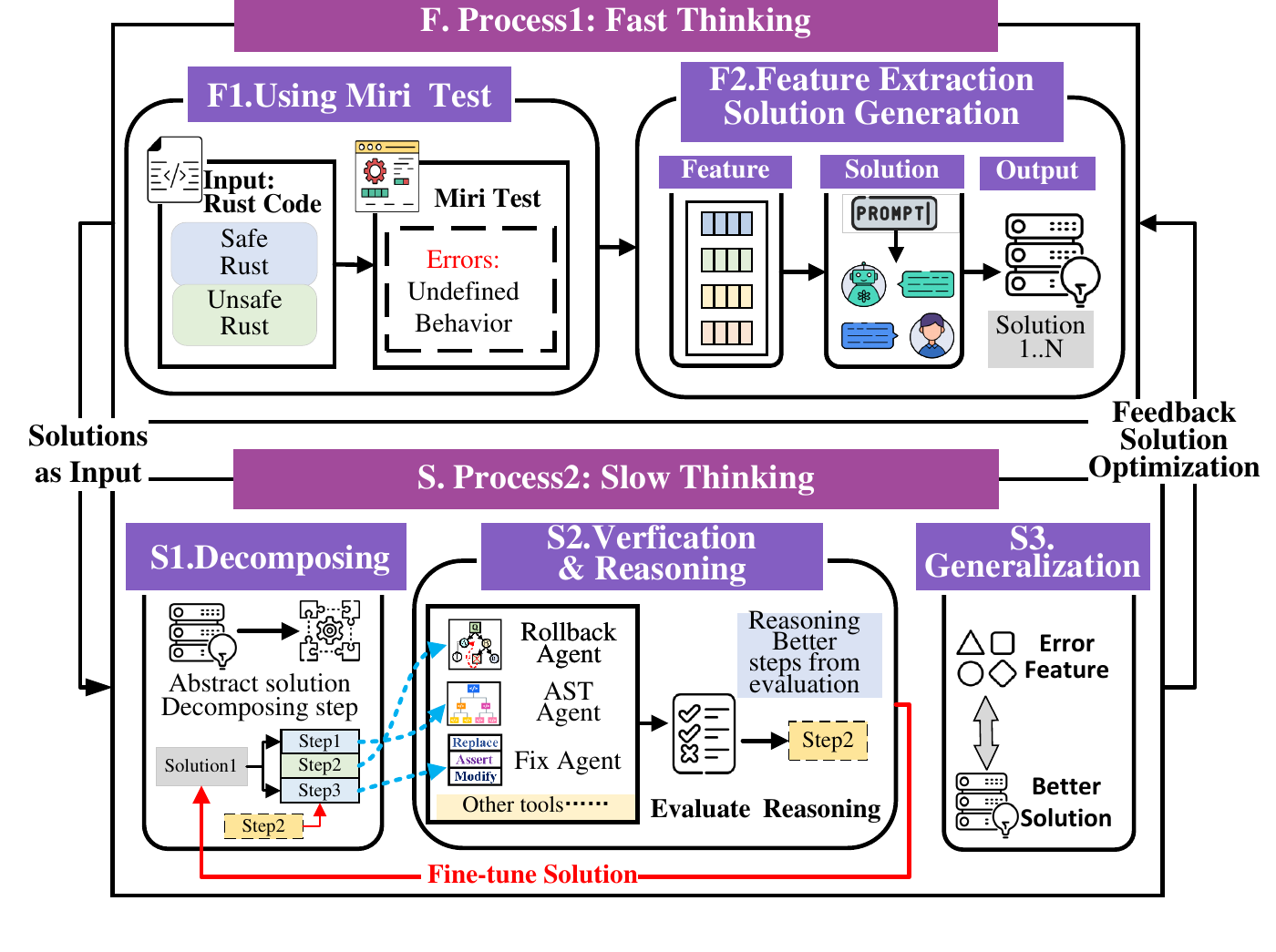}}
\vspace{-1.0em}
\caption{RustBrain Overview: it leverages Miri for static analysis (F1). Extracts and combines features for solution generation (F2). Abstraction and decomposition into steps (S1), integrates multi-agent verification to reduce UBs. Evaluate and reason to fine-tune solution(S2). Generalization provides a solution for such UBs(S3).}
\vspace{-14pt}
\label{part2}
\end{figure}

According to Kahneman's theory~\cite{b14}, human decision-making relies on two main processes: fast thinking, which handles intuitive, quick, often unconscious decisions, and slow thinking, which tackles complex situations through logical reasoning and careful analysis for well-considered choices. 
These are suitable for solving highly specialized problems that require comprehensive analysis and precise judgment. 
We observe that unsafe operations in Rust are limited and rely on intuitive, pattern-driven processes, similar to fast thinking. However, Rust's strong type system requires specialized verification and improvements, similar to slow thinking. 
Therefore, RustBrain is designed as a dual-phase thinking framework incorporating fast and slow thinking (see Fig.~\ref{part2}). 
Fast thinking is intuitive and rapidly processes tasks, generating potential repair solutions based on code features which include error identification, code feature extraction. 
Slow thinking provides deliberate and in-depth processing.  
It combines decomposition, validation, reasoning and generalization to better understand the essence of the problem, delivering more targeted and comprehensive solutions for these types of UBs. 
By leveraging these two types of thinking, the framework achieves adaptive and high-precision repairs, improving both efficiency and reliability in eliminating UBs.  
RustBrain framework involves the following stages: 
\begin{itemize}
    \item \textbf{Stage F1 and F2:} 
    During fast thinking process, Rust code is used as input, and Miri is utilized to detect UBs. 
    If no UBs are found, the process terminates, outputting ``pass''. 
    Conversely, relevant errors, code, and design hints are sent to a LLM agent for feature extraction. Subsequently, these features are integrated with another LLM to rapidly generate multiple repair solutions. 
    \item \textbf{Stage S1:} 
    These solutions are decomposed, validated, reasoned and generalized during the slow thinking process.
    Each solution is broken down into refined steps to enable distributed processing and resolution by different agents. 
    The order of these steps reflects diverse repair strategies, enabling flexible process combinations to optimize repair paths and outcomes.
    \item \textbf{Stage S2:} 
    This stage integrates multiple agents to verify solutions, including agents designed in previous work~\cite{b27,b41}, and new agents specifically developed based on Rust's unique characteristics. 
    These agents include: 
    (i) safe-replacement, assertion-adding, and code-modification agents fix UBs to improve code safety.
    (ii) adaptive rollback agent to reduce hallucinations and propagation in LLMs by reverting to the optimal code state;
    (iii) abstract reasoning agent supports deep semantic understanding and provides more professional fixes. 
    Additionally, the evaluation mechanism is designed to conduct multidimensional assessments of solutions and further refine them through reasoning.
    For instance, increasing the iteration count, adjusting parameters in the repair steps, optimizing the execution path, etc. 
    \item \textbf{Stage S3:} 
    To provide precise and efficient solutions for similar errors, a generalized approach is essential.  
    In the S3 stage, we design a feedback mechanism to match and analyze error features with optimized solutions, combining in-depth evaluation of slow thinking with generation of fast thinking.
    This self-learning approach leverages the strengths of both stages, reducing the dependency of fixed frameworks on knowledge base and enables dynamic solution refinement. 
\end{itemize}

Overall, RustBrain integrates fast and slow thinking to eliminate Rust-UBs, overcoming the limitations of traditional repair frameworks.  
It also combines decomposition, verification, reasoning, and generalization to achieve deep semantic understanding and expert-level support, providing more precise solutions for similar errors while reducing reliance on knowledge bases. 

\section{\textbf{RustBrain Implementation}}
This section explores the technical implementation of the ``Fast and Slow Thinking'' framework.
First, we classify five types of unsafe Rust operations and design the fast thinking process for generating solutions (Section ~\ref{Stage1}). 
Then, we present the slow thinking process to decompose solutions and execute them with multi-agents:  agents for errors fixing with LLMs (Section ~\ref{Stage2}); agents of rollback mechanisms and optimal code selection (Section ~\ref{Stage3}) to mitigate hallucinations; and abstract reasoning agents leveraging Rust semantics for expert guidance(Section ~\ref{Stage4}).
Finally, the results from the slow-thinking are feedback into the fast-thinking to refine evaluations and generate more specialized solutions (Section ~\ref{Stage5}). 

\subsection{Classification of Unsafe Rust and Fast Thinking Process}
\label{Stage1}
To match the error types as closely as possible with the solutions generated by the fast thinking process, we further classify Unsafe Rust from the perspective of error repair.  
This classification allows for targeted problem-solving by identifying the unique characteristics and challenges of each type of unsafe operation.  

\subsubsection{\textbf{Classification of Unsafe Rust}}
Rust's unique design ensures that programs written in Safe Rust have no memory safety issues~\cite{b5,b22}. 
However, Unsafe Rust allows programmers to bypass these safety checks and perform certain operations. 
Unlike C, where unsafe operations are everywhere and are barely checked by the compiler~\cite{b23}, we observe that Rust explicitly classifies unsafe operations and limits them to ``unsafe'' code blocks. 
Even errors arising from interactions between unsafe and safe code occur within the ``unsafe'' block. 
This design significantly reduces the surface area for potential memory safety bugs, thereby facilitating error localization. 

\begin{figure}[htbp]
\centerline{\includegraphics[width=0.4\textwidth, trim=0cm 11cm 23cm 0cm, clip]{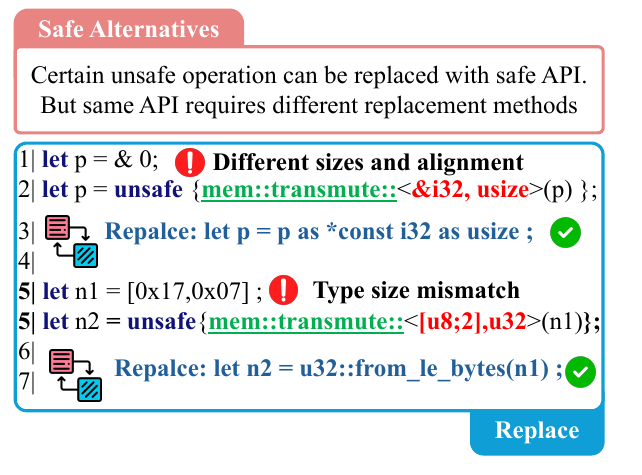}}
\vspace{-1.0em}
\caption{The same API requires different substitution methods.}
\vspace{-1.0em}
\label{listing1}
\end{figure}

Unsafe Rust can be divided into five categories~\cite{b1}: dereference raw pointers, call unsafe functions, implement unsafe traits, access/modify mutable static variables, and access union fields. 
Based on experience and research~\cite{b4,b9}, we further categorize Unsafe Rust from the perspective of error fixing as follows: 
\textbf{(i) Operations with Safe Alternatives:} Certain unsafe operations can be replaced with safer APIs from Rust's standard library. 
\textbf{(ii) Operations Checking via Assertion:} Adding assertions helps detect boundary and type violations, enhancing safety without fundamentally altering the program’s structure.
\textbf{(iii) Operations Semantic Modification:}  
Understanding the underlying intentions of the original code, modifying erroneous semantics, and ensuring that these intentions are preserved in the modifications.
Given Rust's strong type system, we observe that similar unsafe operations often require distinct handling methods. 
For instance, in case (i), 
as shown in Fig~\ref{listing1}, the same unsafe API may require different replacement APIs depending on the context. 
\vspace{-1.0em}

Therefore, based on this classification, the following principles are proposed in this paper to guide the framework design: 
\begin{itemize}
    \item \textbf{Principle 1:} 
    All Unsafe Rust operations are explicitly marked with the ``unsafe'' keyword, reducing the surface area for potential memory safety vulnerabilities facilitates error localization.
    \item  \textbf{Principle 2:}
    The limited types of UBs caused by Unsafe Rust can be further classified into three categories: replaced by Safe Rust, prevented by assertions and require semantic modification.
    \item \textbf{Principle 3:} 
    Under Rust's strong typing characteristics, a single piece of code may exhibit multiple patterns, necessitating that changes be combined with semantic analysis.  
\end{itemize}

\subsubsection{\textbf{Fast Thinking Process}}
Undefined behaviors usually occur at runtime. 
We believe that \textit{prevention is better than cure}, therefore RustBrain integrates tools such as Miri compiler to analyze the Rust code. 
This static analysis occurs at compile time, where the compiler identifies potential UBs within the code, and records them for further examination. 
Once the detection phase is complete, the framework enters the fast thinking process. 

In this process, RustBrain quickly identifies and generates fix solutions based on intuitive Unsafe Rust error pattern recognition. 
Firstly, RustBrain, based on five types of unsafe operations, employs a general-knowledge LLM to rapidly analyze code features, leveraging its generalization capability to identify typical error patterns and guide solution generation. 
According to \textbf{Principle 1}, UBs are generally confined to unsafe regions, which narrows the scope of code analysis and reduces the noise impact of irrelevant code on feature extraction.
Then, since Unsafe Rust operations are limited in types, even with strong typing considerations, similar repair solutions can typically be found for most issues. 
Based on this, RustBrain generates multiple repair solutions tailored to different code characteristics by integrating \textbf{Principle 2}, ensuring that the repair methods are both adaptable and diverse in addressing the unique traits of the code.


\subsection{Slow Thinking Process}
After fast thinking generates preliminary solutions, RustBrain transitions into slow thinking for deeper thinking. 
Due to \textbf{Principle 3}, we need to further decompose, validate, infer, and generalize solutions.
First, RustBrain splits solutions into multiple steps, analyzing and addressing UBs layer by layer. 
Then, it integrates agents design to validate and evaluate each generated solution's compatibility with type and safety requirements. 
We introduce new agents based on Rust features to support deep and professional fixes. 
 
\subsubsection{\textbf{Three Types of Error Fixing Agents}}
\label{Stage2}
\begin{figure}[htbp]
\centerline{\includegraphics[width=0.4\textwidth, trim=0cm 10.7cm 23cm 0cm, clip]{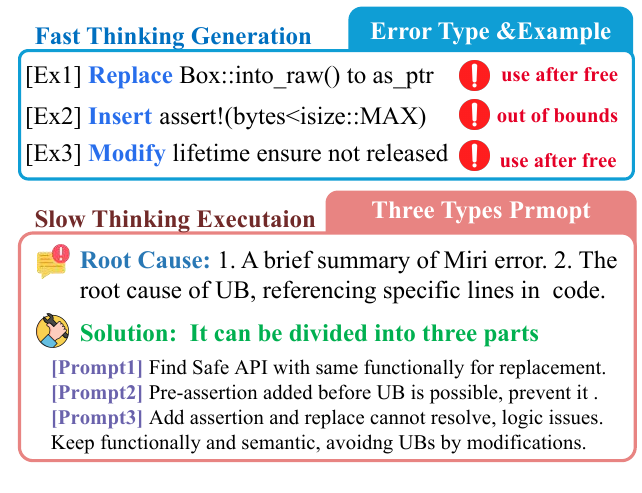}}
\vspace{-1.0em}
\caption{Three types of prompt strategies.}
\vspace{-0.9em}
\label{three_agent}
\end{figure}

To better guide LLM in fixing UBs errors, inspired by \textbf{Principle 2}, RustBrain constructs three types of special agents: equivalent replacement, assertion and modification.
We observe that unsafe Rust is often misused, with some cases where safe code could achieve the same functionality~\cite{b3,b4,b9}. 
Therefore, RustBrain designs an agent to identify situations where safe alternatives can replace unsafe operations. 
This minimizes unsafe code usage, enhancing the overall safety and reliability of Rust. 
Besides, we design an assertion agent that adds assertions in appropriate places to capture potential UBs at runtime before they can propagate. 
This avoids altering the code's semantics while detecting and stopping potential UBs during execution in a timely manner, preventing it from leading to more severe security vulnerabilities.  
However, in certain complex scenarios, UBs involve deeper logical errors or dependencies on external states, and assertions alone cannot directly resolve these issues. 
To address this, RustBrain introduces a modification agent, where adjust the erroneous semantics in the Rust code, ensuring that the code's functionality remains unchanged while effectively reducing the risk of UBs. Prompts are shown in the Fig.~\ref {three_agent}.
\vspace{-1.0em}
\subsubsection{\textbf{Adaptive Rollback and Optimal Code Selection Agent}}
\label{Stage3}
Using multiple fix agents can significantly reduce UBs in Rust code (Section ~\ref{sbsc:Evaluation}). 
However, we observe that in some cases, the number of errors increases after repair, and analyzed and discussed this. 

\begin{figure}[htbp]
\centerline{\includegraphics[width=0.43\textwidth, trim=0.5cm 16.5cm 24cm 0.8cm, clip]{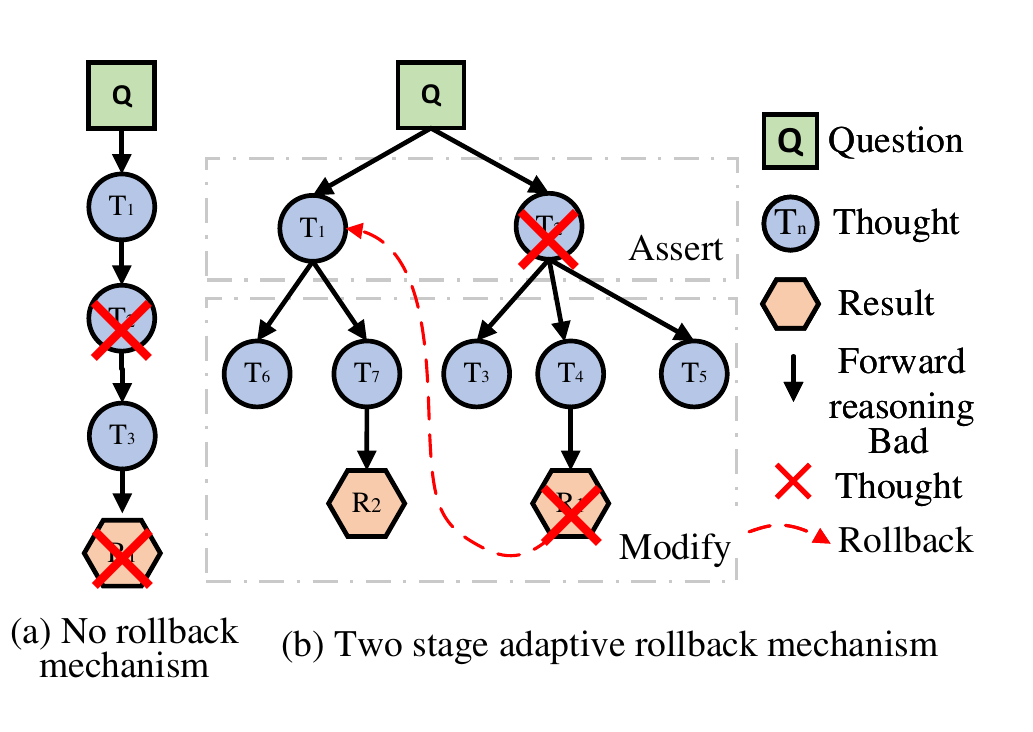}}
\vspace{-1.0em}
\caption{No Rollback Mechanism and Adaptive Rollback Mechanism.}
\vspace{-0.9em}
\label{rollback}
\end{figure}

We define two sequences: \( T = \{T_0, T_1, \dots, T_p\} \) for thoughts generated by slow thinking, and \( N = \{n_0, n_1, \dots, n_p\} \) for errors detected by Miri, where \( T_i \) represents thoughts, \( n_i \) the errors count, and \( p \) the iteration count.
During the slow thinking, the error sequence $ N $ may continue to grow, as in the example sequence $ N_1 = \{1, 3, 4, 6, 9\} $. 
This phenomenon is known as ``large model hallucination" ~\cite{b24,b25}. 
As shown in Fig.~\ref{rollback} (a), without rollback mechanism, the process would proceed from the current error stage to the next stage, exacerbating the impact of hallucinations. 
Existing debugging frameworks~\cite{b26,b27} typically rollback directly to the initial state \( T_0 \) to eliminate accumulated errors and mitigate the propagation of incorrect fixes. 
However, this discards valuable partial corrections made during the iteration process and creates a large overhead ($c*T_{n}$, where \textit{c} is the number of rollbacks)~\cite{b26}. 

We observe another type of error sequence, such as $ N_2 = \{3, 1, 5, 2, 0\} $. 
This indicates that while errors may increase in certain iterations, the overall trend shows a fluctuating decline.
This behavior is similar to humans often providing frequent and random errors during reasoning, but eventually decrease them and converge to a solution.  
Therefore, we suggest that slow thinking a certain degree of self-correcting capability, allowing it to gradually converge towards the correct through multiple iterations. 

\vspace{-2.3em}

Based on the analysis, RustBrain introduces an adaptive rollback mechanism that selects intermediate states intelligently, enabling a more precise and efficient correction process. 
As shown in Fig.~\ref{rollback} (b), before proceeding to the next stage, the process rollback to the optimal code state (the fewest detected errors) to prevent error propagation.
This ensures that subsequent iterations are based on the most refined and stable version of the code, minimizing error propagation and improving overall fix accuracy. 
The rollback overhead is expressed as: $c*T_{n-a}$ (where \textit{a} is the number of thoughts produced in the assertion).
Compared to direct rollback, RustBrain significantly retains the LLM's reasoning capabilities while reducing the overhead generated by reasoning.

\subsubsection{\textbf{Abstract Reasoning Agent}}
\label{Stage4}
To better understand Rust's unique programming semantics and conventions, we propose incorporating a abstract reasoning knowledge agent to facilitate complex relational reasoning and deeper insights. 

Traditional debugging frameworks typically rely on source code snippet libraries, syntax and API specifications, as well as error patterns and repair strategy libraries~\cite{b9,b28,b29} to design knowledge bases. 
However, these methods are not suitable for Rust. 
As explained in \textbf{Principle 3}, Rust’s strong typing characteristics mean that a single unsafe API may exhibit multiple patterns simultaneously, requiring knowledge bases with precise error pattern comprehension and adaptable fixes.  
Existing designs typically rely on pattern matching and learning, which makes it difficult to cover all potential error scenarios~\cite{b9}. 
Additionally, the complex syntax and nested structures in Rust can introduce a large amount of invalid or misleading noise during the construction of the knowledge base, which interferes with the judgment and generation of LLM and limits the effectiveness of these methods. 
Therefore, RustBrain introduces a knowledge base centered on AST (Abstract Syntax Tree) and semantic analysis. 

\vspace{-1.5em}
\begin{figure}[htpb]
\centerline{\includegraphics[width=0.5\textwidth, trim=0.5cm 24cm 20.2cm 0.4cm, clip]{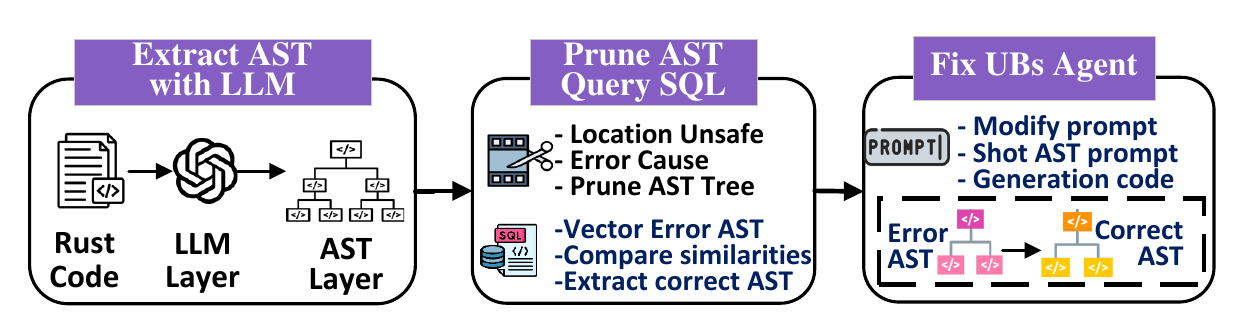}}
\caption{Abstract Reasoning Agent.}
\vspace{-0.6em}
\label{knowledge}
\end{figure}

As shown in Fig.~\ref{knowledge}, RustBrain extracts the code's AST using LLM instead of traditional Rust parsing tools like \textit{syn}. 
The AST constructed by \textit{syn} is overly complex and lacks higher-level semantic understanding. 
LLM not only preserves the semantic context of the code but also flexibly handles the complex contextual relationships.
However, the AST extracted contains irrelevant or noisy information, which distracts the model's attention and decreasing fix performance. 
According to \textbf{Principle 1}, all Unsafe Rust operations are marked with ``unsafe'' keyword.  
RustBrain designs a pruning algorithm (as shown in Algo.\ref{alg:algorithm1}) to retain safety-relevant and functional fix components, minimizing the interference of irrelevant code for the LLM, thus improving the efficiency and accuracy of the repair process. 
\begin{algorithm}[htpb]
	\caption{Pruning irrelevant nodes for Rust AST.}
	\label{alg:algorithm1}
	\KwIn{Original Rust AST: $\mathcal{A}_{orig}$; Miri errors: $\mathcal{E}_{Miri}$;}
	\KwOut{Pruned AST: $\mathcal{A}_{pruned}$;}  
	\BlankLine
	Initialize pruned AST $\mathcal{A}_{pruned} \gets \emptyset$;
	\ForEach{node in $\mathcal{A}_{orig}$}{
		\If{node contains \textnormal{unsafe} keyword}{
			Mark the node as unsafe operation;
			Add node to $\mathcal{A}_{pruned}$;
		}
	}
	\ForEach{unsafe node in $\mathcal{A}_{pruned}$}{
            \If{$\mathcal{E}_{Miri}$ != 0 }{
		      Analyze context of error node;  
        
                \If{context node is irrelevant to unsafe operation}{
				delete context node to $\mathcal{A}_{pruned}$;  
			}
		}
	}
	\Return{$\mathcal{A}_{pruned}$};
\end{algorithm}

Finally, RustBrain vectorizes the pruned AST and uses similarity-based search to query the knowledge base, retrieving repair solutions for error-prone AST structures. These solutions are added to prompts to guide the LLM in generating precise fixes for semantically similar code, which improves repair accuracy (Section~\ref{sbsc:Evaluation}).

\subsection{Feedback Mechanism in Fast and Slow Thinking}
\label{Stage5} 
Fast and slow thinking improves the flexibility and accuracy in eliminating Rust UBs. 
However, the quality of the results from slow thinking are not evaluated against fast thinking solutions.
It limits dynamic adjustment of repair solution based on discovered issues, leading to the persistence of incorrect or inadequate repair solutions and affecting the overall effectiveness of the repair process. 

RustBrain designs a feedback mechanism that integrates validation results into solution generation, allowing further adjustments to initial solutions. 
Specifically, we establish a triplet: \textit{(accuracy, acceptability, overhead)}, to evaluate solutions from multiple dimensions. 
\textit{Accuracy} ensures the  solution passes Miri; \textit{acceptability} checks semantics consistency; and \textit{overhead} measures execution time and resource use.
Through multi-dimensional evaluation and feedback mechanism, fast-thinking is able to absorb the precise analysis and validation information from slow-thinking, allowing the generated repair solutions to achieve a better balance between speed and accuracy, effectively enhancing the overall quality and reliability of the repairs. 
Moreover, this self-learning process enhances RustBrain's generalization ability, supports similar errors repairs, reduces dependency on the knowledge base, and avoids the limitations posed by highly specialized issues.

\section{Evaluation}
\label{sbsc:Evaluation}
The evaluation of RustBrain designed to answer the following research questions: 
\begin{enumerate}
    \item \textbf{RQ1 (Flexibility):} How does RustBrain flexibility affect repair \textit{pass} rate and performance? 
    \item \textbf{RQ2 (Accuracy):} What extent is RustBrain successful in fixing Rust UBs errors as measured by metrics of \textit{pass} and \textit{exec}? 
    \item \textbf{RQ3 (Sensitivity):} How does LLM configurations in GPT-4 impact the performance of RustBrain?
    \item \textbf{RQ4 (Advancement):} 
    How does the effectiveness and performance of RustBrain in eliminating UBs compared to the state-of-the-art method and human expert?
\end{enumerate}

Firstly, we give a brief explanation of the datasets, large language models, baselines and metrics used in this paper.

\noindent{\textbf{Datasets.}}
We use a dataset collected from the official Miri~\cite{b17}, containing code with UBs. The dataset is categorized into different types, including alloc, dangling pointers, and concurrency issues.

\noindent{\textbf{LLMs.}}
We use four LLMs: GPT-3.5~\cite{b18}, GPT-4~\cite{b19}, GPT-O1~\cite{b20}, and Claude 3.5~\cite{b21}, each with a default temperature of 0.5.

\noindent{\textbf{Baselines.}}
Apart from the Miri test, the comparison includes state-of-the-art approaches such as human expert repair (Thetis-Lathe, 2023)~\cite{b3}, and a LLM-based fix method (RustAssistant, 2024)~\cite{b13}. 

\noindent{\textbf{Metrics.}}
All experiments report \textit{pass} rate(\%) and \textit{execution} rate(\%), among which \textit{pass} rate refers to the percentage of code passing Miri after repair, and \textit{execution} rate refers to the percentage of code passing Miri with acceptable semantics. 

Subsequently, we analyze the results to answer these questions. 

\begin{figure}[htbp]
\vspace{-1.5em}
\centerline{\includegraphics[width=0.43\textwidth, trim=0cm 5cm 4cm 0cm, clip]{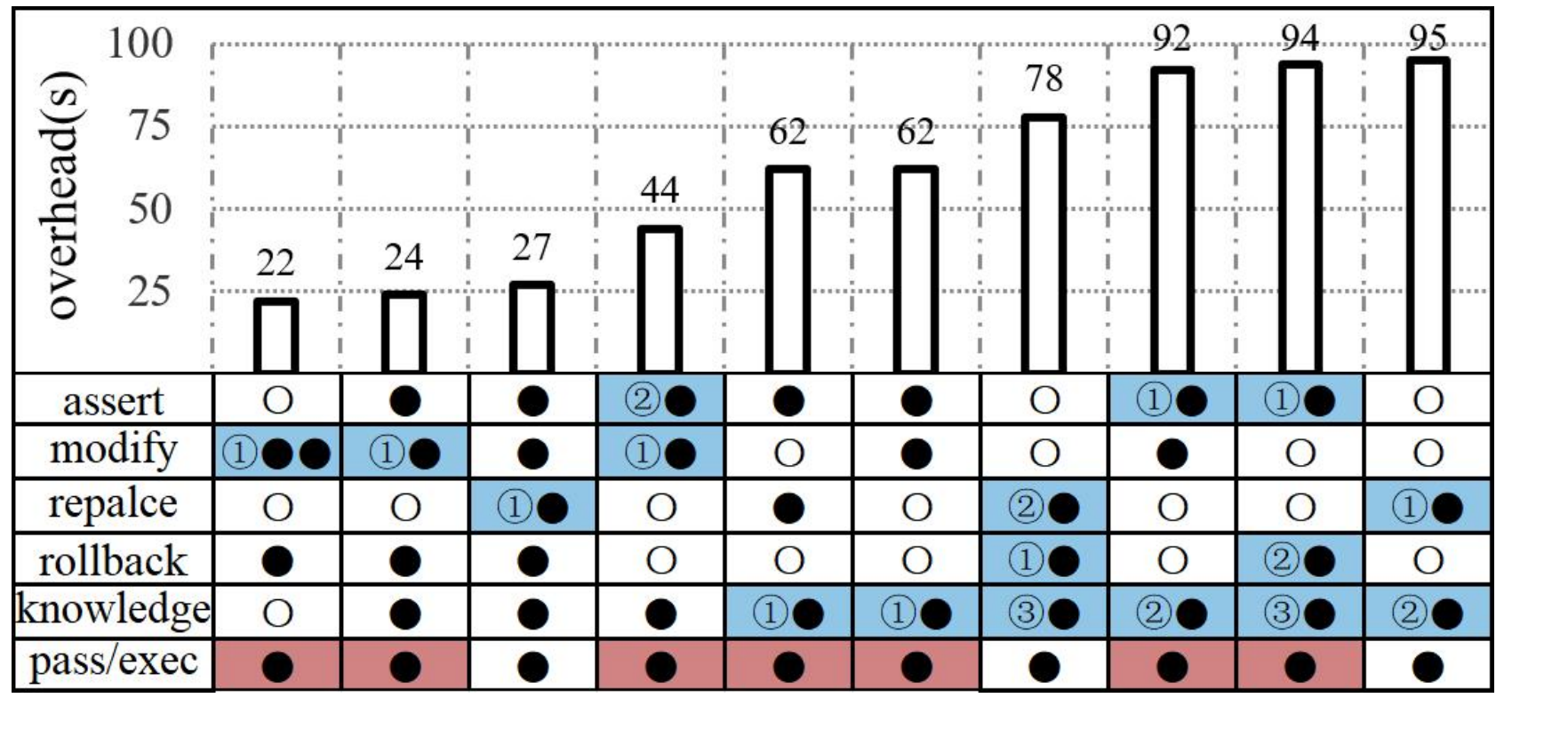}}
\caption{
RustBrain flexibly fixes UBs. Agents, shown on the left, can be enabled [\fullcirc] or disabled [\emptycirc], with serial numbers indicating repair order. Blue denotes agents used in execution, and red indicates semantically acceptable code, otherwise it only passes Miri.
}
\vspace{-2.0em}
\label{flexible}
\end{figure}

\textbf{RQ1 (Flexibility):} 
We select a type of UB (requiring semantic modification) and input it into RustBrain. Ten solutions are generated through fast thinking, followed by reasoning and validation using slow thinking, as depicted in Fig.~\ref{flexible}. 

\begin{figure*}[htpb]
\begin{minipage}[t]{0.5\linewidth}
\centering
\includegraphics[width=\textwidth]{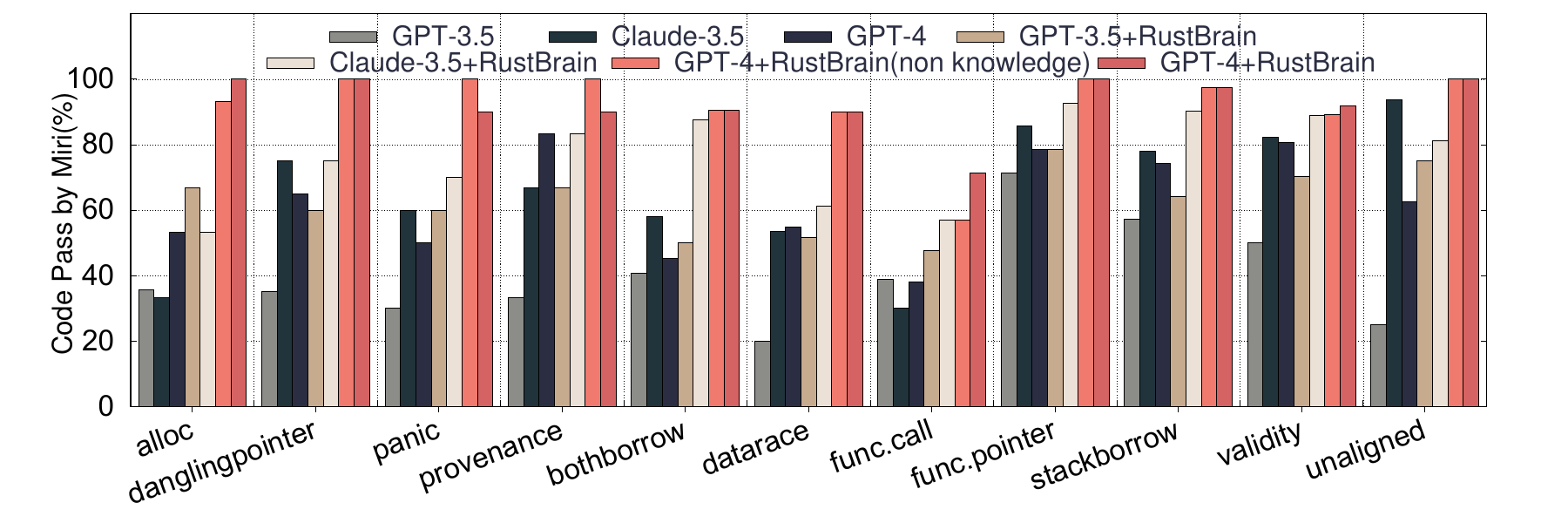}
\vspace{-2.0em}
\caption{RustBrain fixes UBs pass by Miri rate.}
\label{pass1}
\vspace{-1.3em}
\end{minipage}%
\begin{minipage}[t]{0.5\linewidth}
\centering
\includegraphics[width=\textwidth]{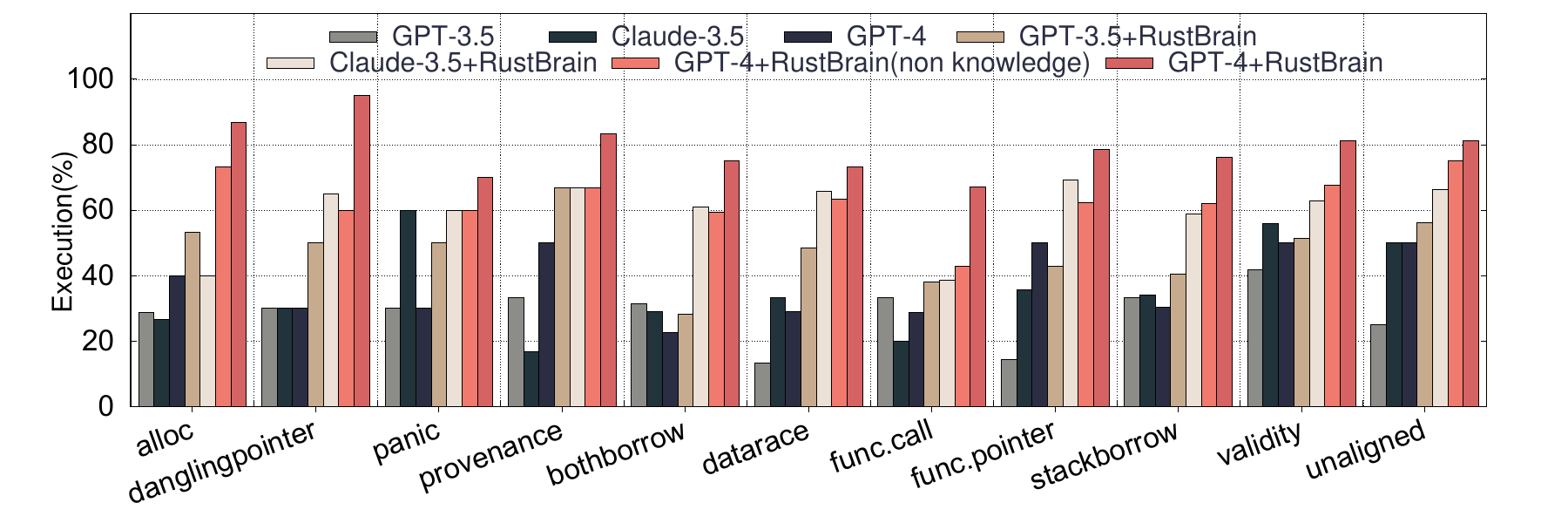}
\vspace{-2.0em}
\caption{RustBrain fixes UBs semantic acceptability rate.}
\label{exec1}
\end{minipage}
\vspace{-1.0em}
\end{figure*}

RustBrain provides more flexible solution and improves the possible and accuracy of fixing UBs. 
The results indicate that: 
\textbf{(i)} For same problem, RustBrain uses fast thinking to generate diverse solutions, overcoming the limitations of fixed-process frameworks that offer only one option, enhancing flexibility and repair effectiveness. 
\textbf{(ii)} Comparing Group 1 with Groups 5, 6, 8, and 9, knowledge base can enhance problem-solving abilities, but introduce 2x-4x overhead, and depends on its size.
Feedback mechanism supports more accurate solution selection, improves repair performance and reduces reliance on knowledge base.  
\textbf{(iii)} Fixed repair framework can cover a wide range of issues, it often includes numerous generic steps, which may introduce unnecessary complexity and overhead  (Group 3). 
RustBrain dynamically designs solutions based on code features, avoiding redundant steps and improve accuracy.
\textbf{(iv)} Specific scenarios exist for the three types of repair solutions for UBs. 
Fixed repair frameworks often provide inappropriate solutions, exacerbate hallucination and reduce accuracy. 
In Groups 3, 7, and 10, even with knowledge base support, semantic acceptable fixes could not be achieved. 
RustBrain, based on code features and feedback mechanism, better guides the generation of solutions for similar error types, avoiding unsuitable fixes and improving both repair accuracy and efficiency.

\textbf{RQ2 (Accuracy):} 
We evaluated RustBrain's repair effectiveness and robustness using various configurations, including different LLMs and knowledge base usage. 
Experimental results, shown in Fig.~\ref{pass1} and Fig.~\ref{exec1}, compare repair \textit{pass} rates and semantic \textit{exec} rates.

Results indicate that for GPT-4 model, integrating RustBrain achieved a \textit{pass} rate of 90.5\% and an \textit{execution} rate of 70.2\%, representing an improvement of 25-35\% compared to using GPT-4 alone for repair. 
Furthermore, when RustBrain is integrated with the knowledge base, it achieves the highest \textit{pass} and \textit{execution} rates, averaging 94.3\% and 80.4\% , respectively.
Additionally, although GPT-3.5 performed lower than GPT-4, combining it with RustBrain achieved a comparable level to GPT-4 in terms of fix \textit{pass} and \textit{execution} rate.
This suggests that the design in RustBrain compensates for the limitations of the base model, improving the accuracy and consistency in handling UBs. 
For Claude-3.5, it exhibits initial semantic capabilities in reducing UBs, comparable to GPT-4.
However, when combined with RustBrain, Claude-3.5 performs less effectively than GPT-4 in understanding complex dependencies and resolving issues.
Compared to standalone Claude-3.5, RustBrain+Claude-3.5 improves the \textit{pass} rate and \textit{execution} rate by 17\% and 20\%, respectively.

\begin{figure}[t]
\vspace{-1.0em}
\centerline{\includegraphics[width=0.45\textwidth]{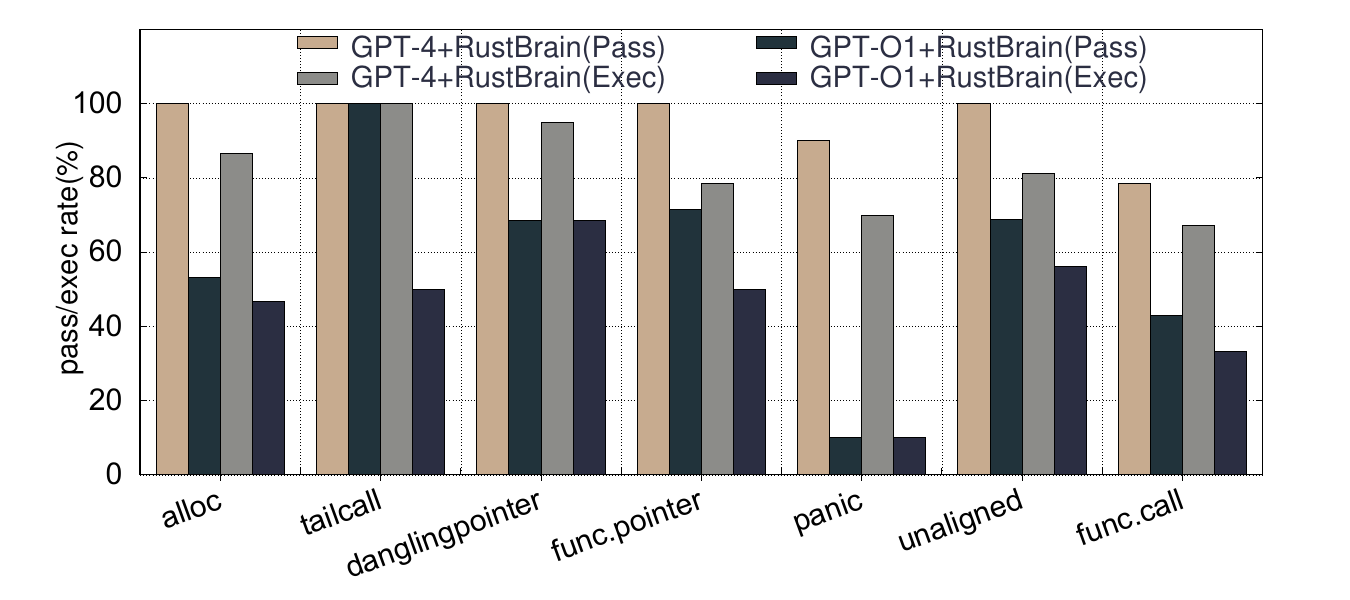}}
\vspace{-1.0em}
\caption{Comparison of RustBrain with GPT-O1 on UBs Repair.}
\label{gpto1}
\end{figure}

Besides, we also compare RustBrain with the state-of-the-art GPT model, GPT O1. 
Due to O1's high cost, a subset of error types was analyzed, with results shown in Fig.~\ref{gpto1}. 
Despite GPT-O1's exceptional reasoning, its repair effectiveness remains inferior to RustBrain. 
For uncommon errors like \textit{panic} errors, GPT-O1 fails to provide suitable solutions based on code features. 
In contrast, RustBrain, leverages the combination of fast and slow thinking to generate more targeted solutions, improving \textit{execution} rate by 35.6\%.  

\textbf{RQ3 (Sensitivity):} 
Temperature is a key parameter affecting output diversity in model generation~\cite{b30}. 
We conduct sampling tests in Miri dataset and evaluate the stability of the repair results by calculating the confidence intervals~\cite{b31}. 
The Confidence Interval(CI) reflects the uncertainty of fixing UBs in LLMs, ensuring that the reported pass and \textit{exec} rates are reliable within a certain probability range (c=95\%). 
In the experiments, we analyze the impact of different temperature settings on the performance of RustBrain in GPT-4.

\begin{figure}[t]
\vspace{-1.3em}
\centerline{\includegraphics[width=0.45\textwidth]{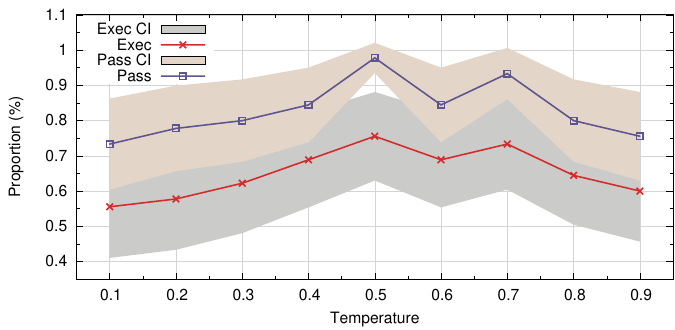}}
\vspace{-0.8em}
\caption{Different Temperatures on RustBrain Repair Efficacy.}
\label{temp}
\vspace{-1.5em}
\end{figure}

As shown in Fig.~\ref{temp}, \textit{pass} and \textit{exec} rates reach highest points of 97\% and 77\% with a temperature of 0.5. 
Higher temperatures increase the flexibility of outputs. However, this flexibility comes at the cost of semantic integrity, leading to a decrease in semantic acceptability (e.g, 0.7). In contrast, lower temperatures limit flexibility and potentially missing opportunities to repair UBs.

\begin{figure}[t]
\vspace{-1.0em}
\centerline{\includegraphics[width=0.5\textwidth]{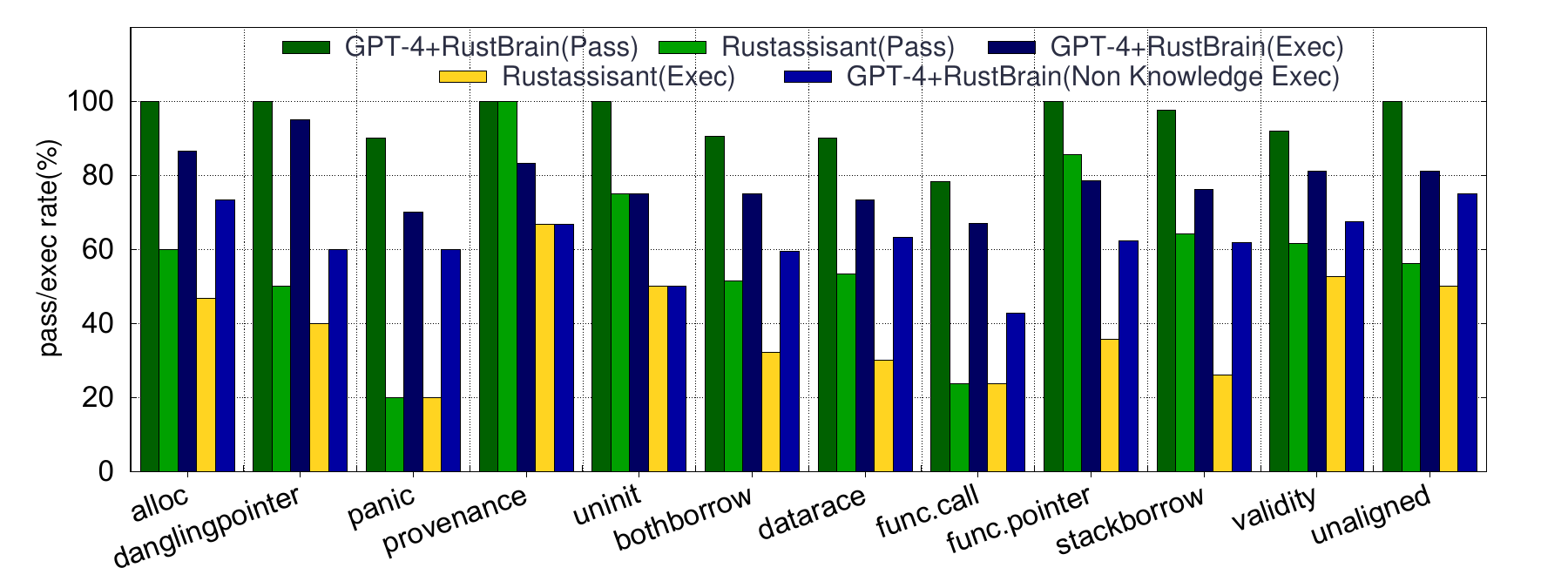}}
\vspace{-0.8em}
\caption{Comparison of RustBrain with RustAssisant on UBs Repair.}
\label{rustassisant}
\vspace{-1.5em}
\end{figure}

\textbf{RQ4 (Advancement): } 
To evaluate the effectiveness and efficiency of RustBrain, compared to RustAssistant (the state-of-the-art Rust fix tool) and human experts. 
As shown in Fig.~\ref{rustassisant}, compared to RustAssisant, the \textit{pass} rate and \textit{execution} rate of RustBrain increased by 33\% and 41\%, respectively. 
This indicates that RustBrain offers greater flexibility, breaking the limitations of existing repair frameworks. It is able to address issues that the current frameworks cannot solve, thereby reducing UBs and improving overall safety.

To evaluate RustBrain's effectiveness compared to human experts, we assessed their performance across various error types, as shown in Table~\ref{human}. 
While human experts excel in debugging, the framework demonstrates competitive performance in addressing UBs caused by complex semantic rules. 
On average, RustBrain achieves a 7x speedup across different error types, with the gap widening for more complex or rare errors. 
For example, in \textit{function pointer} UBs, which mainly involve type conversion problems and require deeper Rust expertise for guidance, RustBrain shows up to 18x. 
Furthermore, we note that when the dataset contains similar UBs, RustBrain can reduce dependency on the knowledge base and generate precise solutions through the feedback mechanism, effectively minimizing UBs and overhead, as shown in the red sections of Table~\ref{human}. 

Overall, fixed frameworks like RustAssistant rely on their knowledge base for speedup over human experts, but their problem-solving ability is constrained by the knowledge they contain. 
RustBrain leverages a feedback mechanism that feeds back evaluated solutions from the slow-thinking phase into the fast-thinking phase. 
Through this self-learning process, the fast-thinking can generate more precise solutions, without being fully dependent on the design of the knowledge base, reducing the overhead and time to fix UBs.

\begin{table}[htpb]
    \renewcommand{\arraystretch}{1.1}
    \caption{Execution Time of RustBrain against Human}
    \label{human}
    \centering
    \resizebox{0.98\linewidth}{!}{%
    \begin{tabular}{|c|c|c|c|c|}
        \hline
        \multirow{3}{*}{\textbf{Types}} & \multicolumn{3}{c|}{\textbf{Average Time/s}} &\multirow{3}{*}{\textbf{Speedup}}\\ 
        \cline{2-4} 
         &  \multicolumn{2}{c|}{\textbf{GPT-4+RustBrain}}   &  \multirow{2}{*}{\textbf{Human}}    &    \\ 
         \cline{2-3} 
         & No\_knowledge & knowledge & &\\
        \hline
        stack borrow & 86 & 123& 366 & 2.81x     \\  
        unaligned pointer & 74 & 118 & 222 & 4.63x     \\ 
        \rowcolor{green!20}
         \textbf{validity} & \textbf{79} & \textbf{109} & \textbf{678} & \textbf{8.58x}     \\  
          \rowcolor{green!20}
         \textbf{alloc} & \textbf{40} & \textbf{77} & \textbf{450} & \textbf{11.25x} \\ 
         \rowcolor{green!20}
         \textbf{func. pointer} & \textbf{39} & \textbf{50}& \textbf{480} & \textbf{12.3x }    \\
         provenance & 45 & 64 & 240 & 5.33x     \\  
         panic & 76 & 107 & 336& 4.42x     \\  
          \rowcolor{red!20}
         \textbf{func. calls} & \textbf{65} & \textbf{74} & \textbf{1176}& \textbf{18.1x}     \\ 
         \rowcolor{red!20} 
         \textbf{dangling pointer} & \textbf{70} & \textbf{74} & \textbf{114} & \textbf{1.63x}     \\ 
         \rowcolor{red!20}
         \textbf{both borrow} & \textbf{69} & \textbf{87} & \textbf{762} & \textbf{11x}      \\ 
         concurrency & 45 & 45 & 144 & 3.2x      \\ 
         data race & 63 & 91 & 336& 5.3x     \\ \hline
         \textbf{Average} & \textbf{62.6} & \textbf{84.9} &  \textbf{442}& \textbf{7.4x}\\
        \hline
    \end{tabular}}
\vspace{-1.0em}
\end{table}

\section{Related Work}
Efforts to improve Rust's safety target Unsafe Rust, divided into manual, semi-automated, and fully automated methods.
Manual methods, like formal methods~\cite{b33,b34}, static analysis~\cite{b35,b36}, rely on developer expertise for precise fixes but are time-consuming and error-prone. 
Semi-automated methods, such as symbolic execution~\cite{b37,b38}, combine tool-based detection with developer analysis. 
Fully automated methods use machine learning~\cite{b9,b39} and LLMs~\cite{b12,b13} to reduce human involvement but face limitations in accuracy, coverage, and contextual understanding. 
Our approach employs a collaboration of fast and slow thinking to provide flexible and precise repair solutions while maintaining the safety of the Rust. 

\section{Conclusion}
We propose RustBrain, a dynamic, automated framework to eliminate UBs in Rust projects.
Combine fast and slow thinking two phases, RustBrain overcomes traditional framework limitations, offering flexible fixes and enhancing Rust code safety. 
The fast-thinking phase extracts code features to generate multiple fixes, while the slow-thinking phase employs Rust feature-based agents to execute solutions. 
Feedback from execution refines decision-making. 
Experimental show RustBrain achieves \textit{pass} rate of 94.3\% and \textit{exec} rate 80.4\%, with up to 7x speedup compared with human experts. 
Future work, we are interested in providing automated safety enhancements for complex Rust code involving multi-module calls.




\vspace{12pt}
\color{red}

\end{document}